\documentclass{Odyssey2026}

\usepackage{numprint}
\npdecimalsign{.}
\npthousandthpartsep{}
\newcommand{\eer}[1]{\nprounddigits{2}\numprint{#1}}
\newcommand{\dcf}[1]{\nprounddigits{3}\numprint{#1}}

\usepackage{float}
\usepackage{multirow}
\usepackage{booktabs}

 \interspeechcameraready

\title{Kiwano: A Cutting-Edge Open-Source Toolkit for Speaker Verification}

\author[affiliation={1}]{Mickael}{Rouvier}
\author[affiliation={1}]{Pierre Michel}{Bousquet}

\affiliation{Avignon University}{LIA, UPR 4128}{France}
\email{firstname.lastname@univ-avignon.fr}

\usepackage{comment}
\usepackage{soul}
\begin{document}

\maketitle

\begin{abstract}
In this paper, we present Kiwano, an open-source toolkit designed to advance research and evaluation for speaker verification. Kiwano provides a lightweight yet extensible framework built on PyTorch, offering standardized recipes, pretrained models, and integration of several widely used speaker verification architectures. The toolkit emphasizes reproducibility, by delivering transparent training pipelines, unified evaluation protocols and ready-to-use baselines across multiple corpora. Beyond conventional training and inference, Kiwano includes tools for benchmarking, experiment tracking and rapid prototyping of new architectures. To foster community adoption, the toolkit is distributed under the Apache 2.0 license, accompanied by comprehensive documentation and reproducible experiments. By lowering entry barriers and standardizing evaluation practices, Kiwano contributes a valuable resource for both academic research and applied development in speaker verification. The toolkit is publicly available at: \url{https://github.com/kiwano-toolkit/kiwano/}
\end{abstract}

\section{Introduction}
Speaker embeddings~\cite{snyder2018x,thienpondt2023ecapa2,peng2023attention} have become the standard representation for speaker identity for tasks such as Speaker Verification (SV), diarization and speech processing adaptations. These compact fixed-dimensional vectors capture speaker-specific traits and are scored using back-ends like cosine similarity or Probabilistic Linear Discriminant Analysis (PLDA) for decision-making. Beyond traditional SV, speaker embeddings also play a crucial role in downstream applications such as text-to-speech synthesis and voice conversion.

As research on speaker embeddings has matured, the need for accessible and reproducible toolkits has become increasingly important. The speech community has actively contributed to open-source frameworks, from early toolkits like ALIZE~\cite{larcher2013alize} or Kaldi~\cite{povey2011kaldi} to modern PyTorch-based toolkits such as SpeechBrain~\cite{ravanelli2021speechbrain}, ESPnet~\cite{jung2024espnet} and WeSpeaker~\cite{wang2023wespeaker}. These have democratized research, but gaps persist: many prioritize general speech tasks over SV-specific needs, lack up-to-date architectures or overlook production features like efficient back-ends and domain adaptation. Evaluation campaigns such as VoxCeleb Speaker Recognition Challenge~\cite{huh2024vox} or NIST SRE~\cite{sre2006nist} highlight innovative techniques; however, replicating State-Of-The-Art (SOTA) results often requires custom engineering that is not available in existing implementations.


Despite the rapid progress of speaker verification systems and the availability of several open-source toolkits, important questions remain insufficiently explored. In particular, the impact of training dynamics, architectural scaling and reproducibility on modern speaker verification performance is still not well understood. Reported improvements are often obtained under different training settings, making fair comparisons difficult and limiting insights into what truly drives performance gains. Furthermore, computational aspects such as scalability, efficiency and training stability are rarely analyzed jointly with accuracy. As a result, there is a growing need for unified experimental frameworks that enable systematic and reproducible analysis of these factors. In this work, we address this need through Kiwano, a unified open-source framework for speaker verification, and through a controlled empirical study of three key factors: global mini-batch scaling, network depth, and run-to-run reproducibility.


Kiwano is a research and production-oriented toolkit for learning high-quality speaker embeddings. Named after the resilient kiwano fruit, symbolizing robustness in diverse environments, Kiwano is purely PyTorch-based. Its key advantages include:

\begin{figure}[t]
    \centering
    \includegraphics[width=0.30\textwidth]{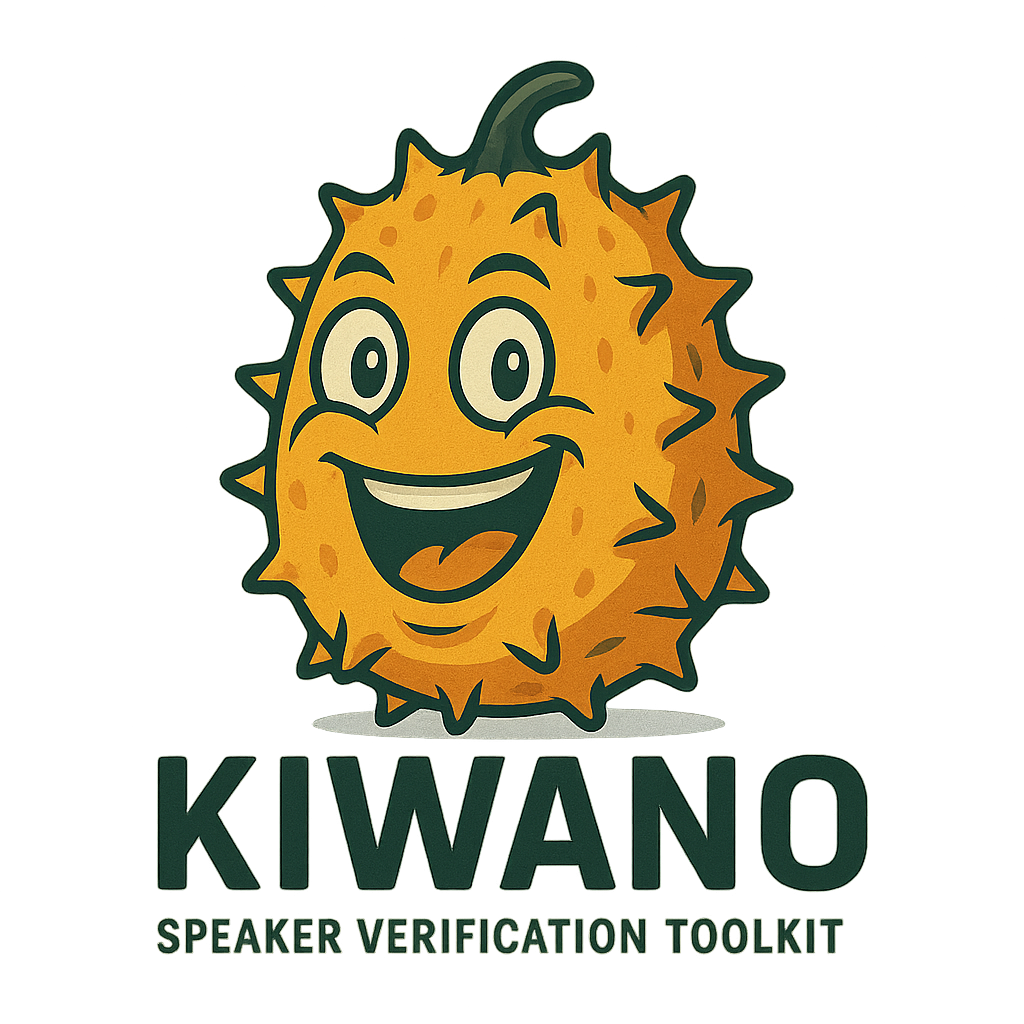}
    \caption{Kiwano is an open-source framework for speaker verification research and evaluation.}
    \label{fig:example}
\end{figure}

\begin{itemize}


\item \textbf{Up-to-Date Models}: Kiwano integrates the latest architectures, such as Xi-Vector~\cite{lee2021xi} a Bayesian extension of the traditional speaker embedding; ECAPA2~\cite{thienpondt2023ecapa2} a hybrid neural network architecture that combines 1D and 2D convolutions; and ReDimNet~\cite{yakovlev24_interspeech} designed to improve representational efficiency through optimized scaling of depth and feature dimensionality.

\item \textbf{Advanced Back-Ends}: In contrast to many other toolkits, Kiwano offers a suite of back-end tools, encompassing scoring methods (including cosine similarity and PLDA), normalization techniques (including Symmetric Normalization : S-Norm~\cite{shum2010unsupervised}, Adaptive Score Normalization : AS-Norm~\cite{karam2011towards}, and Adaptive Data Normalization - AD-Norm~\cite{cumani2023adaptive}) and domain adaptation techniques (including CORAL~\cite{Alam18}, CORAL+~\cite{Lee18} and fDA~\cite{Bousquet2019}) to manage domain mismatches.

\item \textbf{Extensive Recipes}: Kiwano provides an extensive collection of reproducible recipes for datasets including VoxCeleb~\cite{chung2018voxceleb2}, CN-Celeb~\cite{fan2020cn,li2022cn}, CommonBench~\cite{hintz2024commonbench}, VoxTube~\cite{yakovlev2023voxtube}, VoxBlink~\cite{lin2024voxblink}, DiPCo~\cite{rouvier2022far} and 3D-Speaker~\cite{zheng20233d} covering comprehensive data downloading and preparation processes.

\end{itemize}


Kiwano achieves competitive performance, with Equal Error Rates (EER) reaching 0.34\% on the VoxCeleb1-O benchmark. Kiwano is designed as a unified experimental framework enabling systematic and reproducible investigation of modern speaker verification systems. The toolkit is publicly available here~\footnote{\url{https://github.com/kiwano-toolkit/kiwano/}}.






The paper is organized as follows:
Section~\ref{sec:related_work} reviews existing open-source toolkits and related efforts in speaker verification.
Section~\ref{sec:system_components} details the main components of the toolkit, including the front-end, back-end and data management modules.
Section~\ref{sec:experimental_results} reports experimental results across various models and back-end scoring methods and compares Kiwano with other toolkits.
Finally, Section~\ref{sec:conclusions} presents our conclusions.

\section{Related work}
\label{sec:related_work}

The development of open-source toolkits has played a central role in the progress of speaker verification research. Early frameworks such as Kaldi~\cite{povey2011kaldi} provided comprehensive infrastructures for data preparation, feature extraction and model training. Initially built around the i-vector paradigm~\cite{dehak2010front} and later extended to x-vector approaches~\cite{snyder2018x}, Kaldi remains widely used within the community. However, its C++-based architecture is less suited to rapid prototyping and modern deep learning workflows, which has motivated the emergence of more flexible PyTorch-based toolkits.

A new generation of speaker verification toolkits has therefore been developed using PyTorch, offering improved accessibility and faster experimentation. SpeechBrain~\cite{ravanelli2021speechbrain} provides a general-purpose speech processing framework supporting speaker verification alongside other tasks such as automatic speech recognition, text-to-speech and speaker diarization. It includes implementations of ECAPA-TDNN and ResNet models and provides recipes for VoxCeleb. While highly versatile, its multi-task orientation can introduce additional complexity for speaker verification-focused experimentation.

WeSpeaker~\cite{wang2023wespeaker} adopts a production-oriented perspective and focuses specifically on speaker embedding learning. Built on PyTorch, it offers scalable data management through Unified IO (UIO) and supports architectures such as TDNN, ResNet and Self-Supervised Learning (SSL) based systems. It also provides recipes for major datasets including VoxCeleb, CN-Celeb and VoxConverse~\cite{chung20_interspeech}. ESPnet-SPK~\cite{jung2024espnet} extends the ESPnet framework with a complete speaker verification pipeline and reproducible recipes. It integrates self-supervised front-ends such as WavLM and supports a wide range of architectures, while maintaining interoperability with other ESPnet modules, enabling multi-task and multimodal experimentation.

These toolkits have significantly lowered the entry barrier to speaker verification research and improved reproducibility across studies. Nevertheless, they differ in scope and design objectives: some emphasize research flexibility and multi-task learning, whereas others focus on scalability and production deployment.


In contrast, Kiwano is designed not only as a speaker verification toolkit but also as a unified experimental framework enabling reproducible analysis of modern speaker verification systems. It integrates recent SOTA models, including ECAPA2, ReDimNet and Xi-Vector, and provides reproducible recipes across several widely used datasets such as VoxCeleb, CN-Celeb, CommonBench and others. Beyond front-end modeling, Kiwano incorporates a comprehensive set of back-end tools, including cosine scoring, Linear Discriminant Analysis (LDA) and PLDA, enabling fully reproducible end-to-end experimentation and controlled comparison of architectural and training choices.

\section{System components}
\label{sec:system_components}


Kiwano is designed as a modular and extensible framework for end-to-end speaker verification. Its architecture can be divided into three main components: data management (Section~\ref{sec:data_management}), front-end speaker embedding module (Section~\ref{sec:front_end}) and  back-end scoring module (Section~\ref{sec:back_end}). These components are coordinated through a recipe layer that ensures reproducibility across datasets and experimental setups.




\subsection{Data Management}
\label{sec:data_management}

The data management component handles input organization and metadata preparation, performs acoustic feature extraction, applies diverse data augmentation techniques and manages resampling and chunking to standardize input lengths.

The input organization relies on a list-based metadata system for dataset management, inspired by practices from large-scale production environments. This design enables efficient handling of datasets ranging from just a few hours to millions of utterances without incurring out-of-memory issues. All relevant information such as utterance paths, speaker identifiers and durations are stored in a list. During training, utterances are opened on demand, rather than preloaded into memory, which both minimizes memory overhead and facilitates parallel data loading.

Feature preparation (including data augmentation, resampling, chunking...) is performed online, removing the need for pre-computed feature storage and reducing disk usage. This on-the-fly processing also improves model robustness by presenting it with slightly different input variations at each training epoch.

Thus, the overall processing pipeline includes the following steps:

\begin{itemize}
\item \textbf{Speaker Mapping} : Maps speaker names to IDs.
\item \textbf{Augmentation}: Supports both signal-level methods (e.g., noise addition, reverberation, speed perturbation, air absorption, codec simulation, aliasing, clipping distortion and sign flip) and feature-level techniques (e.g., SpecAugment, which introduces masking while correcting mean shifts between training and test data).
\item \textbf{Feature Extraction}: Computes traditional acoustic features such as log Mel-filter banks (FBank) or spectrograms.
\item \textbf{Resampling} and \textbf{Chunking}: Adjusts sample rates and cuts utterances to fixed lengths (e.g., 350 frames during training), with padding for short segments.
\item \textbf{Normalization}: Applies Cepstral Mean and/or Variance Normalization (CMVN) per utterance.
\end{itemize}

\subsection{Front-end: Speaker Embedding}
\label{sec:front_end}

The front-end module comprises various encoder networks, a pooling layer and classification head, which together convert variable-length utterances into compact, fixed-dimensional speaker embeddings that capture speaker-specific traits.

For the encoder network, the toolkit implements a wide range of SOTA architectures, allowing researchers to explore various modeling paradigms for speaker representation learning. The main models available are:

\begin{itemize}

\item \textbf{fwSE-ResNet-200}: Built on a 200-layer ResNet, it is organized in a (4, 30, 30, 4) block configuration. Each residual block contains three 2D convolutional layers activated by the SiLU (Sigmoid Linear Unit) function, which enhances gradient flow and nonlinearity. Feature Squeeze-and-Excitation (fwSE) modules are incorporated only in the first and second residual stages, allowing effective channel-wise feature recalibration in the early layers while maintaining computational efficiency in deeper layers~\cite{rouvier2021studying}. The network is trained with the Additive Margin Softmax (AM-Softmax) loss and optimized using the Jeffreys loss~\cite{bousquet2023jeffreys}, which promotes a balanced trade-off between intra-class compactness and inter-class separability.


\item \textbf{ReDimNet}~\cite{Yakovlev_2024}: A flexible model capable of alternating between 1D and 2D processing of speech features. It does this by reshaping data between 1D and 2D forms, allowing the model to take advantage of both temporal analysis (common in 1D models) and spectral–spatial processing (typical of 2D models). This reshaping makes it possible to combine outputs from 1D and 2D layers using simple residual connections, without losing information or changing the overall data volume. As a result, ReDimNet can efficiently extract speaker embeddings while maintaining a good balance between accuracy and computational cost, scaling smoothly from lightweight to larger configurations depending on available resources.

\item \textbf{ECAPA2}~\cite{thienpondt2023ecapa2}: A hybrid neural network that combines 1D and 2D convolutional architectures. ECAPA2 integrates two complementary components: a Local Feature Extractor, which captures fine-grained and spatially invariant patterns across narrow frequency regions and a Global Feature Extractor, which aggregates local features across the full frequency spectrum. This dual-path structure balances local detail sensitivity with global contextual consistency. The architecture follows the official implementation from the ECAPA2 paper and HuggingFace, trained with Sub-Center Additive Angular Margin (AAM) Softmax loss and optimized using the Jeffreys loss, replacing the original margin-mixup strategy to achieve improved performance.

\item \textbf{Xi-Vector}~\cite{lee2021xi}: A Bayesian extension of the traditional x-vector framework, Xi-Vector enhances speaker embedding extraction by incorporating uncertainty estimation. Built upon the fwSE-ResNet-200 backbone, it adds an auxiliary branch that estimates frame-level uncertainty, allowing the network to assign greater importance to reliable frames and less to noisy or ambiguous ones. This uncertainty is integrated through a linear Gaussian model at the pooling stage, effectively combining the generative modeling principles of i-vectors with the discriminative learning strengths of x-vectors.

\end{itemize}

Different pooling strategies are implemented in Kiwano (including statistics pooling, attentive statistics pooling and multi-head attention pooling). The toolkit also supports various margin-based loss functions such as Additive Angular Margin Softmax (AAM-Softmax), AM-Softmax and Sub-Center loss, complemented by sub-center learning and inter-top-k penalties to further enhance robustness. Together, these components allow users to explore a wide range of embedding extraction paradigms within a unified framework.

\subsection{Back-end: Scoring and Calibration}
\label{sec:back_end}

In contrast to many toolkits that focus mainly on embedding extraction, Kiwano also provides a comprehensive set of back-end modules for speaker verification, including:

\begin{itemize}
    \item \textbf{Scoring methods}: Kiwano supports both cosine similarity and PLDA.
    
    \item \textbf{Embedding pre-processing}: Before scoring, embeddings can be processed using centering, whitening, length normalization, and LDA for dimensionality reduction.
    
    \item \textbf{Domain adaptation}: To improve robustness under domain mismatch between training and evaluation conditions, Kiwano integrates several adaptation techniques: CORAL, CORAL+, and feature Distribution Adaptor (fDA), which align covariance statistics across domains.
    
    \item \textbf{Score normalization}: Kiwano includes several score normalization strategies, such as S-Norm, AS-Norm, and D-Norm.

    \item \textbf{Score calibration and quality-aware normalization}: Kiwano also supports post-processing strategies based on the Consistency Measure Factor (CMF)~\cite{zheng2024score} and Quality Measure Functions (QMF)~\cite{thienpondt2021idlab}. These methods further refine verification scores by accounting for score consistency and trial-dependent quality information.

\end{itemize}





\begin{table*}[t]
\resizebox{\textwidth}{!}{  
\centering
\setlength\tabcolsep{2.4pt}
\begin{tabular}{l c c c c c c c c c c c c c c c c c c c}
\toprule

\multirow{3}{*}{\textbf{Model}} 
& \multicolumn{5}{c}{\textbf{Computational cost}} 
& \multicolumn{6}{c}{\textbf{In-domain}} 
& \multicolumn{6}{c}{\textbf{Out-of-domain}}
& \multicolumn{2}{c}{\textbf{Global}}
\\

\cmidrule(lr){2-6}
\cmidrule(lr){7-12} 
\cmidrule(lr){13-18}
\cmidrule(lr){19-20}

& \textbf{Train} & \textbf{\#Params} & \textbf{\#GPU} & \textbf{GPU} & \textbf{Energy}
& \multicolumn{2}{c}{\textbf{VoxCeleb1-O}} 
& \multicolumn{2}{c}{\textbf{VoxCeleb1-E}} 
& \multicolumn{2}{c}{\textbf{VoxCeleb1-H}} 
& \multicolumn{2}{c}{\textbf{CN-Celeb}} 
& \multicolumn{2}{c}{\textbf{DiPCo}} 
& \multicolumn{2}{c}{\textbf{CommonBench}}
& \multicolumn{2}{c}{\textbf{Average}} \\

\cmidrule(lr){7-8} \cmidrule(lr){9-10} \cmidrule(lr){11-12}
\cmidrule(lr){13-14} \cmidrule(lr){15-16} \cmidrule(lr){17-18} \cmidrule(lr){19-20}

& \textbf{(hours)}  &  & \textbf{used} & & \textbf{(kWh)}
& \textbf{EER}$\downarrow$ & \textbf{minDCF}$\downarrow$ 
& \textbf{EER}$\downarrow$ & \textbf{minDCF}$\downarrow$ 
& \textbf{EER}$\downarrow$ & \textbf{minDCF}$\downarrow$ 
& \textbf{EER}$\downarrow$ & \textbf{minDCF}$\downarrow$ 
& \textbf{EER}$\downarrow$ & \textbf{minDCF}$\downarrow$ 
& \textbf{EER}$\downarrow$ & \textbf{minDCF}$\downarrow$
& \textbf{EER}$\downarrow$ & \textbf{minDCF}$\downarrow$  \\

\midrule

fwSE-ResNet-200  
& 79h & 83M & 32 & 82\% & 706.3 
 & \eer{0.4998537834998987} & \dcf{0.052488536741195795} 
& \eer{0.6164003340544839} & \dcf{0.05528196165406397} 
& \eer{1.0938583708503167} & \dcf{0.09636452291611082} 
& \eer{10.272874323374372} & \dcf{0.4557731697586416} 
& \eer{3.6467999999999985} & \dcf{0.20985999999999994} 
& \eer{2.8193476842470284} & \dcf{0.2906647569367406}
& \eer{3.158189082671017} & \dcf{0.1934054913344588} \\

ECAPA2 
& 316h & 30M & 32 & 85\%  
& 2835.9 & \eer{0.6567222951798217} & \dcf{0.04318297584682351} 
& \eer{0.6691755010447572} & \dcf{0.07261811441361654} 
& \eer{1.1873427548808868} & \dcf{0.11115666977174142} 
& \eer{14.677484997250195} & \dcf{0.5466599189899334} 
& \eer{6.1533999999999995} & \dcf{0.46366533333333343} 
& \eer{3.595719871216492} & \dcf{0.326993794199483}
& \eer{4.4899742366} & \dcf{0.2607128011}  \\

Xi-Vector 
& 79h & 83M & 32 & 82\% 
& 706.3 & \eer{0.5158065638243636} & \dcf{0.05009713391511416} 
& \eer{0.6667609502677933} & \dcf{0.06280489849582821} 
& \eer{1.1601142856429936} & \dcf{0.10496445053144057} 
& \eer{12.120563508033372} & \dcf{0.49588235496058314} 
& \eer{3.713933333333331} & \dcf{0.22480666666666663} 
& \eer{2.9767833048188637} & \dcf{0.2980820115602757}
& \eer{3.5256603243} & \dcf{0.2061062527} \\

ReDimNet-B6 
& 65h & 15M & 32 & 86\% &  
  536.1  & \eer{0.5769583935574966} & \dcf{0.06243428719327536} 
& \eer{0.7360930511492352} & \dcf{0.07788173030918051} 
& \eer{1.3518027090777676} & \dcf{0.12373894300447014} 
& \eer{12.650051999712408} & \dcf{0.5372616467156005} 
& \eer{5.086599999999997} & \dcf{0.2942866666666666} 
& \eer{4.132194562364513} & \dcf{0.3621658178065059}
& \eer{4.088950119310236} & \dcf{0.2429615152826165} \\

\bottomrule
\end{tabular}
}

\caption{
Comparison of different architectures across in-domain and out-of-domain evaluation sets together with computational cost. 
Training time, model size, number of GPUs, GPU memory usage and energy consumption are reported to highlight the trade-off between performance and efficiency.
}
\label{tbl:overall_results}
\end{table*}

\subsection{Training Strategies}

Kiwano leverages PyTorch and provides several advanced functionalities to facilitate efficient optimization and large-scale experimentation. The key components are outlined below:

\begin{itemize}

\item \textbf{Learning Rate and Margin Schedule}: The learning rate scheduler implemented in Kiwano follows a three-phase design: warmup, plateau, and decay. Each phase is fully configurable to adapt to different training setups. During the warmup phase (first 5 epochs), the learning rate increases linearly from 1e-5 to 0.1 while the margin is kept at 0, allowing the model to stabilize early training dynamics. In the plateau phase (next 25 epochs), the learning rate remains fixed at 0.1 while the margin is linearly increased from 0 to 0.2 to progressively enforce larger angular separation between speakers. Finally, in the decay phase, after the margin reaches its maximum, the learning rate decreases exponentially by a factor of 0.5 every 10 epochs, promoting convergence and fine-grained optimization. L2 weight regularization is set to 1e-4,

\item \textbf{Large Margin Fine-Tuning}: An additional stage with larger margins (e.g., 0.5 in AAM-Softmax or AM-Softmax) and longer input segments (e.g., 5 seconds), providing consistent performance improvement,

\item \textbf{Distributed Training}: Enables efficient multi-node and multi-GPU training through PyTorch’s DistributedDataParallel module and HuggingFace’s Accelerate toolkit, ensuring seamless scalability across diverse hardware configurations and full compatibility with Slurm-based job scheduling environments.


\end{itemize}

\section{Experimental Results}
\label{sec:experimental_results}


In this section, we first describe the datasets and experimental setup (Section~\ref{sec:dataset_setup}). We then evaluate several speaker embedding architectures within the Kiwano framework under both in-domain and out-of-domain conditions (Section~\ref{sec:embedding_architectures}). 

Next, we investigate the impact of several practical factors, namely training dynamics, architectural scaling, and reproducibility. More specifically, we analyze the influence of the global mini-batch size (Section~\ref{impact_mini_batch_size}), network depth (Section~\ref{impact_network_depth_architectures}), and repeated training on performance reproducibility (Section~\ref{impact_repeated_training_performance}). We also study the effect of additional refinement techniques, including model averaging, Large-Margin Fine-Tuning (LM-FT), CMF, AS-Norm, and QMF (Section~\ref{sec:impact_lmft_asnorm}). Finally, we compare Kiwano with other open-source speaker verification toolkits (Section~\ref{sec:comparison_with_other_toolkits}).


\subsection{Datasets and Setup}
\label{sec:dataset_setup}

\noindent \textbf{Datasets}: We use VoxCeleb2-dev (5,994 speakers, ~1M utterances) for training. Model performance is evaluated on both in-domain and out-of-domain benchmarks. The VoxCeleb1-O/E/H serve as in-domain evaluations, assessing performance under conditions similar to the training data. To measure generalization and cross-domain robustness, we further evaluate on out-of-domain datasets: CN-Celeb, DiPCo and CommonBench. CN-Celeb~\cite{li2022cn} is a large-scale, real-world Chinese speaker recognition dataset designed to capture cross-domain and multilingual variability. DiPCo~\cite{rouvier2022far} is a meeting-style dataset featuring multi-speaker conversations recorded using distant microphones. CommonBench~\cite{hintz2024commonbench} is a recently introduced multilingual evaluation benchmark that aggregates recordings from various open-source speech datasets.

\hspace{1mm}


\noindent \textbf{Models and Training}: Front-ends include fwSE-ResNet-200, ECAPA2, ReDimNet and Xi-Vector. All embeddings are 256-dimensional, trained for 51 epochs with a batch size of 512 and evaluated using cosine back-ends. The AM-Softmax loss (margin = 0.2, scale = 30) is used for all models except ECAPA2, which employs AAM-Softmax.

The MUSAN dataset~\cite{snyder2015musan} is used to generate additive noise, while simulated Room Impulse Responses (RIRs) are applied for reverberation. For each utterance in the training set, either additive noise or reverberation is applied (but not both simultaneously). Speed perturbation is also performed by varying the playback speed by a factor of 0.9 or 1.1; the resulting audio is treated as originating from a new speaker due to pitch shift effects.

Acoustic features are extracted according to the settings reported in the original paper for each model. Most systems use 80-dimensional log Mel-filterbank (FBank) features with a 25 ms window and a 10 ms frame shift. However, to remain consistent with their original best-performing configurations, ReDimNet uses 72-dimensional log Mel-filterbank features, while ECAPA2 uses 256-dimensional spectrogram features. All training data are segmented into 350-frame chunks, followed by cepstral mean normalization without variance normalization. In addition, SpecAugment~\cite{Park_2019} is applied to improve robustness against overfitting and channel variability.

\hspace{1mm}

\noindent\textbf{Hardware and computational cost}: All experiments were conducted on the Jean Zay supercomputer. Unless otherwise stated, experiments were performed on NVIDIA Tesla V100 GPUs with 32 GB of memory. The only exception is Section~\ref{impact_network_depth_architectures}, where NVIDIA H100 GPUs with 80 GB of memory were used because of the higher memory requirements of very deep networks. Energy consumption, CPU usage, and GPU usage were monitored using the Compute Energy and Emissions Monitoring Stack (CEEMS)~\cite{paipuri2024ceems}, an open-source platform-independent tool designed to measure the energy usage and emissions of individual workloads on high-performance computing and cloud infrastructures. CEEMS collects fine-grained power measurements directly from node-level hardware sensors.

\hspace{1mm}

\noindent \textbf{Evaluation metrics}: We report the EER and minimum Detection Cost Function (minDCF) as primary evaluation metrics with P$_{target}$=0.01 and the cost of miss and false alarm set to 1.

\subsection{Embedding Architectures}
\label{sec:embedding_architectures}

Table~\ref{tbl:overall_results} reports the results obtained with different embedding architectures integrated into the Kiwano framework. Performance is evaluated on both in-domain (VoxCeleb1-O/E/H) and out-of-domain (CN-Celeb, DiPCo, and CommonBench) test sets. All results are reported without large-margin fine-tuning (LM-FT) or AS-Norm.

Among all systems, fwSE-ResNet-200 provides the best trade-off between performance and computational cost, achieving a global average of 3.16\% EER and 0.193 minDCF while maintaining a moderate training cost (79 hours, 706.3 kWh). It also delivers the strongest in-domain performance, obtaining the best results on VoxCeleb1-O, VoxCeleb1-E, and VoxCeleb1-H. Under out-of-domain conditions, fwSE-ResNet-200 remains the most robust system overall, yielding the best average performance across CN-Celeb, DiPCo, and CommonBench.

ECAPA2 and ReDimNet-B6 remain competitive on the in-domain benchmarks, but both exhibit substantial performance degradation on the out-of-domain datasets. ECAPA2, for instance, achieves 0.66\% EER on VoxCeleb1-O but degrades to 14.68\% EER on CN-Celeb, resulting in the weakest global performance (4.49\% EER and 0.261 minDCF). It is also the most computationally expensive model, requiring 316 hours of training and 2835.9 kWh of energy. In contrast, ReDimNet-B6 is the most efficient architecture in terms of model size (15M parameters), training time (65 hours), and energy consumption (536.1 kWh), but this efficiency comes at the expense of lower overall performance, especially under domain mismatch, with a global average of 4.09\% EER and 0.243 minDCF. Xi-Vector also remains competitive and performs close to fwSE-ResNet-200 on the in-domain benchmarks, but shows lower robustness in out-of-domain conditions, leading to a higher global average (3.53\% EER and 0.206 minDCF) than fwSE-ResNet-200.






\subsection{Impact of mini-batch size}
\label{impact_mini_batch_size}

The mini-batch size plays a critical role during training, influencing both optimization stability and generalization performance. In distributed training, the effective (global) mini-batch size is defined as the product of the number of GPUs and the local mini-batch size processed on each GPU. In this section, we analyze the impact of the global mini-batch size using the fwSE-ResNet-200 architecture. Similar trends were observed across the other architectures integrated in Kiwano (ECAPA2, Xi-Vector, and ReDimNet); therefore, only fwSE-ResNet-200 results are reported for clarity.

Table~\ref{tbl:minibatch} reports the impact of the global mini-batch size on fwSE-ResNet-200. The best overall performance is obtained with a global mini-batch size of 256, using 16 GPUs, which achieves the lowest out-of-domain average (\(5.40\%\) EER, \(0.315\) minDCF), albeit with a much longer training time (150 hours). Increasing the global mini-batch size to 512 significantly reduces training time to 79 hours while maintaining competitive performance. In particular, it yields the best in-domain EER (\(0.74\%\)) and remains close to the 256 setting on out-of-domain evaluations. Larger-scale training with a global mini-batch size of 1024, using 32 GPUs, further reduces training time but consistently degrades performance. These results indicate that Kiwano is efficiently designed for both moderate and large-scale GPU configurations, while a global mini-batch size of 512 offers the best trade-off between training efficiency and verification performance.

\begin{table}[!htbp]
\resizebox{\columnwidth}{!}{  
\centering
\setlength\tabcolsep{3pt}
\begin{tabular}{l c c c c c c}
\toprule

\multirow{2}{*}{\textbf{System}} 
& \textbf{Train}
& \textbf{Mini-batch}
& \multicolumn{2}{c}{\textbf{In-domain Avg}} 
& \multicolumn{2}{c}{\textbf{Out-of-domain Avg}} \\

\cmidrule(lr){4-5}
\cmidrule(lr){6-7}

& \textbf{(hours)}
& \textbf{size} 
& \textbf{EER}$\downarrow$ & \textbf{minDCF}$\downarrow$
& \textbf{EER}$\downarrow$ & \textbf{minDCF}$\downarrow$ \\

\midrule

fwSE-ResNet-200
& 150h 
& 256
& \eer{0.74938614} & \textbf{\dcf{0.06478292956}}
& \textbf{\eer{5.3981047407}} & \textbf{\dcf{0.315432777}} \\

fwSE-ResNet-200
& 109h
& 384
& \eer{0.7650942533} & \dcf{0.06962091121}
& \eer{5.9313802192} & \dcf{0.3350755738} \\

fwSE-ResNet-200
& 79h 
& 512
& \textbf{\eer{0.7367041628015663}} & \dcf{0.06804500710379019}
& \eer{5.579674002540466} & \dcf{0.31876597556512737} \\

fwSE-ResNet-200
& 47h 
& 1024
& \eer{0.7972225742} & \dcf{0.07440881627}
& \eer{5.7565905358} & \dcf{0.331060174} \\

\bottomrule
\end{tabular}
}
\caption{
Impact of global mini-batch size on fwSE-ResNet-200 performance. A mini-batch size of 256 yields the best overall verification results, while a size of 512 provides a better trade-off between performance and training time.
}
\label{tbl:minibatch}
\end{table}

\subsection{Impact of Network Depth}
\label{impact_network_depth_architectures}

Model depth is an important factor in speaker embedding extraction, as deeper architectures may capture more diverse and discriminative speaker characteristics. At the same time, increasing depth also raises computational cost and does not necessarily translate into better performance once sufficient model capacity has been reached. In this section, we investigate the impact of ResNet depth on speaker verification performance.

We evaluate four fwSE-ResNet configurations with increasing depth, namely 100, 200, 400, and 600 layers. All models share the same architectural design and training protocol, and differ only in the number of residual blocks.

Table~\ref{tbl:depth_resnet} reports the averaged results over in-domain datasets (VoxCeleb1-O/E/H) and out-of-domain datasets (CN-Celeb, DiPCo, and CommonBench). The results show that shallower networks provide the best performance on the in-domain benchmarks. In particular, fwSE-ResNet-100 achieves the lowest in-domain average with 0.73\% EER and 0.065 minDCF. By contrast, deeper architectures yield more competitive results under domain mismatch. fwSE-ResNet-400, for example, achieves the best out-of-domain average EER (5.48\%), while fwSE-ResNet-200 provides the best out-of-domain minDCF (0.319), confirming that increased depth can improve robustness in more challenging conditions.

However, increasing the depth up to 600 layers does not lead to further improvements. Although fwSE-ResNet-600 remains competitive, it does not outperform the 200- or 400-layer configurations, while significantly increasing training cost and energy consumption. This suggests that the benefit of increasing depth saturates beyond a certain point.

\begin{table}[!htbp]
\resizebox{\columnwidth}{!}{  
\centering
\setlength\tabcolsep{3pt}
\begin{tabular}{l c c c c c c}
\toprule

\multirow{2}{*}{\textbf{System}} 
& \textbf{Train}
& \textbf{\#Params}
& \multicolumn{2}{c}{\textbf{In-domain Avg}} 
& \multicolumn{2}{c}{\textbf{Out-of-domain Avg}} \\

\cmidrule(lr){4-5}
\cmidrule(lr){6-7}

& \textbf{(hours)}  & 
& \textbf{EER}$\downarrow$ & \textbf{minDCF}$\downarrow$
& \textbf{EER}$\downarrow$ & \textbf{minDCF}$\downarrow$ \\

\midrule

fwSE-ResNet-100
& 24h & 250.0 &
 \textbf{\eer{0.7339907884}} & \textbf{\dcf{0.06530638052}}
& \eer{5.7489785676} & \dcf{0.335844975} \\

fwSE-ResNet-200
& 36h & 370.9 & 
\eer{0.7367041628015663} & \dcf{0.06804500710379019}
& \eer{5.579674002540466} & \textbf{\dcf{0.31876597556512737}} \\

fwSE-ResNet-400
& 65h & 653.5 & \eer{0.7765163741} & \dcf{0.07136686525}
& \textbf{\eer{5.4843398176}} & \dcf{0.3194262481} \\

fwSE-ResNet-600
& 74h & 1149.3 & \eer{0.7658648333} & \dcf{0.06480069599} 
& \eer{5.605815082} & \dcf{0.3265764095} \\

\bottomrule
\end{tabular}
    }
\caption{
Impact of ResNet depth on speaker verification performance. Results are averaged across in-domain and out-of-domain evaluation sets. fwSE-ResNet-200 provides the best trade-off between performance and model complexity.
}
\label{tbl:depth_resnet}
\end{table}

\subsection{Impact of repeated training on performance reproducibility}
\label{impact_repeated_training_performance}

Training deep speaker verification systems involves several stochastic components, including random parameter initialization, data shuffling, and on-the-fly augmentation. As a result, retraining the same model with identical hyperparameters may lead to slight performance variations. Assessing this variability is important to evaluate the reproducibility and reliability of the proposed training pipeline. In this section, we therefore analyze the impact of repeated training on performance consistency.

To this end, we train the same fwSE-ResNet-200 model four times under strictly identical conditions. All runs use the same training data (VoxCeleb2), architecture, optimization procedure, augmentation pipeline, and hyperparameters. The only differences arise from stochastic initialization and sample ordering during training.

Table~\ref{tbl:repro} reports the results obtained across the four independent runs. Overall, the variability remains limited for both in-domain and out-of-domain evaluations. On the in-domain benchmarks, the average performance is \(0.7725\%\) EER and \(0.0695\) minDCF, with standard deviations of only \(0.022\) and \(0.0019\), respectively. On the out-of-domain benchmarks, the mean performance reaches \(5.7125\%\) EER and \(0.3265\) minDCF, with standard deviations of \(0.167\) and \(0.0057\). The corresponding coefficients of variation remain low, ranging from \(0.017\) to \(0.029\), which indicates that the dispersion across runs is small relative to the average performance.

These results confirm that the proposed training pipeline is highly reproducible. Although minor variations are observed between runs, they remain limited in magnitude and do not affect the overall conclusions. Hence, the performance differences reported in the other experiments can be attributed primarily to architectural or training changes rather than to random fluctuations inherent to stochastic optimization.

\begin{table}[!htbp]
\resizebox{\columnwidth}{!}{  
\centering
\setlength\tabcolsep{3pt}
\begin{tabular}{l c c c c}
\toprule

\multirow{2}{*}{\textbf{System}} 
& \multicolumn{2}{c}{\textbf{In-domain Avg}} 
& \multicolumn{2}{c}{\textbf{Out-of-domain Avg}} \\

\cmidrule(lr){2-3}
\cmidrule(lr){4-5}

& \textbf{EER}$\downarrow$ & \textbf{minDCF}$\downarrow$
& \textbf{EER}$\downarrow$ & \textbf{minDCF}$\downarrow$ \\

\midrule

fwSE-ResNet-200 (run~1)
& \eer{0.7367041628015663} & \dcf{0.06804500710379019}
& \eer{5.579674002540466} & \dcf{0.31876597556512737} \\

fwSE-ResNet-200 (run~2)
& \eer{0.7723507443} & \dcf{0.07256996436}
& \eer{5.6933418697} & \dcf{0.329887093} \\

fwSE-ResNet-200 (run~3)
& \eer{0.7837486554} & \dcf{0.06819919805}
& \eer{5.9836490751} & \dcf{0.3320124478} \\

fwSE-ResNet-200 (run~4)
& \eer{0.7971742989} & \dcf{0.06929229209}
& \eer{5.5933677595} & \dcf{0.3252125486} \\

\midrule
\textbf{Mean}
& \textbf{\eer{0.7725}} & \textbf{\dcf{0.0695}}
& \textbf{\eer{5.7125}} & \textbf{\dcf{0.3265}} \\

\textbf{Std}
& 0.022 & 0.0019
& 0.167 & 0.0057 \\

\textbf{Coeff. Var (std/mean)}
& 0.028 & 0.027
& 0.029 & 0.017 \\

\bottomrule
\end{tabular}
}
\caption{
Impact of repeated training on performance reproducibility using fwSE-ResNet-200. Mean, standard deviation (std) and coefficient of variation (std/mean) are reported across multiple independent runs. Low dispersion confirms the stability and reproducibility of the proposed training pipeline.
}
\label{tbl:repro}
\end{table}

\subsection{Impact of Average models, LM-FT, CMF, AS-Norm and QMF}
\label{sec:impact_lmft_asnorm}

We evaluate five techniques to improve performance beyond the base embedding extractor: model averaging, large-margin fine-tuning (LM-FT), consistency-aware score calibration (CMF), Adaptive Score Normalization (AS-Norm), and Quality Measure Function (QMF) calibration.

Model averaging uses the last 10 checkpoints to reduce optimization noise and stabilize inference. LM-FT further improves class separability through a larger angular margin and longer training segments. At inference time, CMF rescales scores according to their consistency, AS-Norm normalizes them using an adaptive impostor cohort of 200 speakers from the training set, and QMF calibrates them using the raw similarity score together with the enrollment and test magnitudes, defined as the average similarity scores between the corresponding representation and the impostor cohort.

\begin{table}[h]
\resizebox{\columnwidth}{!}{  
    \centering
    \setlength\tabcolsep{2.6pt}
    \begin{tabular}{l c c c c c c}
    \toprule
    & \multicolumn{2}{c}{\textbf{VoxCeleb1-O}} & \multicolumn{2}{c}{\textbf{VoxCeleb1-E}} & \multicolumn{2}{c}{\textbf{VoxCeleb1-H}} \\
        & \textbf{EER}$\downarrow$ & \textbf{minDCF}$\downarrow$ & \textbf{EER}$\downarrow$ & \textbf{minDCF}$\downarrow$ & \textbf{EER}$\downarrow$ & \textbf{minDCF}$\downarrow$ \\
    \midrule
    fwSE-ResNet-200 & \eer{0.4998537834998987} & \dcf{0.052488536741195795} & \eer{0.6164003340544839} & \dcf{0.05528196165406397} & \eer{1.0938583708503167} & \dcf{0.09636452291611082}   \\

    \hspace{5pt} + Average models & \eer{0.45199544252650353} & \dcf{0.0396766301457292} & \eer{0.6012231291867537} & \dcf{0.0548989856569551}  & \eer{1.074435423146813} & \dcf{0.09299969205771405}  \\

    \hspace{5pt} ++ LM-FT & \eer{0.46263062940948024} & \dcf{0.04105946175938735} & \eer{0.5912200045312963} & \dcf{0.05291591291684858}  & \eer{1.0501112699219046} & \dcf{0.09448294724894508}  \\

    \hspace{5pt} +++ CMF & \eer{0.4333833706373113} & \dcf{0.04546800701031616} & \eer{0.5794922007494329} & \dcf{0.05148092619957813}  & \eer{1.0246980319665362} & \dcf{0.09661477355147792}  \\

    \hspace{5pt} ++++ AS-Norm & \eer{0.3748888530929735} & \dcf{0.04690402445526886} & \eer{0.5496552304461181} & \dcf{0.04899740624744168}  & \eer{0.9680628159517177} & \dcf{0.08859638246541242}  \\    

    \hspace{5pt} +++++ QMF & \eer{0.3376676783784872} & \dcf{0.04184941090011451} & \eer{0.537237540736008} & \dcf{0.0472898466981123}  & \eer{0.9248603114275947} & \dcf{0.08594974121088231}  \\

    \bottomrule
    \end{tabular}
    
}
    \caption{Effect of Average models, LM-FT, CMF, AS-Norm and QMF on the fwSE-ResNet-200 model, evaluated on VoxCeleb1-O/E/H. Both EER and minDCF are reported.}  
    \label{tbl:training_strategies}
\end{table}

Table~\ref{tbl:final_training_strategies} presents the final performance obtained with the complete refinement pipeline, while intermediate results are omitted for brevity. Overall, the full pipeline yields substantial relative gains, reducing the EER by approximately 30\% on VoxCeleb1-O, 11\% on VoxCeleb1-E, and 15\% on VoxCeleb1-H.

\begin{table}[!htbp]
\resizebox{\columnwidth}{!}{  
    \centering
    \setlength\tabcolsep{2.6pt}
    \begin{tabular}{l c c c c c c}
    \toprule
    & \multicolumn{2}{c}{\textbf{VoxCeleb1-O}} & \multicolumn{2}{c}{\textbf{VoxCeleb1-E}} & \multicolumn{2}{c}{\textbf{VoxCeleb1-H}} \\
        & \textbf{EER}$\downarrow$ & \textbf{minDCF}$\downarrow$ & \textbf{EER}$\downarrow$ & \textbf{minDCF}$\downarrow$ & \textbf{EER}$\downarrow$ & \textbf{minDCF}$\downarrow$ \\
    \midrule
    fwSE-ResNet-200  & \textbf{\eer{0.3376676783784872}} & \dcf{0.04184941090011451} & \textbf{\eer{0.537237540736008}} & \textbf{\dcf{0.0472898466981123} } & \textbf{\eer{0.9248603114275947}} & \textbf{\dcf{0.08594974121088231}}  \\

    ECAPA2 & \eer{0.46} & \textbf{\dcf{0.04094917093245276}} & \eer{0.59} & \dcf{0.05580653578467226}  & \eer{1.01} & \dcf{0.091202327047330388}  \\ 


    ReDimNet & \eer{0.41} & \dcf{0.04094917093245276} & \eer{0.66} & \dcf{0.05580653578467226}  & \eer{1.14} & \dcf{0.11002327047330388}  \\

    Xi-Vector & \eer{0.36} & \dcf{0.04194917093245276} & \eer{0.6} & \dcf{0.05080653578467226}  & \eer{0.98} & \dcf{0.09012327047330388}  \\

    \bottomrule
    \end{tabular}
    
}
    \caption{Final performance of the evaluated architectures on VoxCeleb1-O, VoxCeleb1-E, and VoxCeleb1-H after applying the complete post-training and score refinement pipeline, including model averaging, LM-FT, CMF, AS-Norm, and QMF.}
    \label{tbl:final_training_strategies}
\end{table}

\subsection{Comparison with Other Toolkits}
\label{sec:comparison_with_other_toolkits}

We compared Kiwano with other open-source speaker verification toolkits, including WeSpeaker, ESPnet-SPK, and 3D-Speaker, using the best-performing reported model for each framework under comparable experimental settings. To ensure the fairest possible comparison, we only considered systems trained on the same training corpus (VoxCeleb2), and evaluated under the same cosine scoring protocol without applying score normalization methods such as AS-Norm. This choice was motivated by the fact that not all toolkits report results with AS-Norm (or related normalization methods), which would otherwise make the comparison less consistent. The results on the VoxCeleb1 benchmark datasets (O, E, and H) are reported in Table~\ref{tbl:other_toolkit}. We note that ESPnet-SPK does not report results for VoxCeleb1-E and VoxCeleb1-H.

As shown, Kiwano (fwSE-ResNet-200) achieves EERs of 0.46\% on VoxCeleb1-O, 0.64\% on VoxCeleb1-E, and 1.13\% on VoxCeleb1-H, outperforming the best configurations from WeSpeaker (ResNet-293), ESPnet-SPK (SKA-WavLM), and 3D-Speaker (ERes2Net-large-lm). The gains are particularly notable on VoxCeleb1-H, where Kiwano achieves a relative EER reduction of around 13\% compared with the best competing system. These results confirm that Kiwano provides highly competitive performance on standard evaluation benchmarks.



\begin{table}[!htbp]
\resizebox{\columnwidth}{!}{  
    \centering
    \setlength\tabcolsep{2.6pt}
    \begin{tabular}{l c c c}
    \toprule
    & \textbf{VoxCeleb1-O} & \textbf{VoxCeleb1-E} & \textbf{VoxCeleb1-H} \\
        & \textbf{EER}$\downarrow$ & \textbf{EER}$\downarrow$ & \textbf{EER}$\downarrow$ \\
    \midrule
    Kiwano (fwSE-ResNet-200) & \textbf{\eer{0.45731303596799194}} & \textbf{\eer{0.6357167259925183}} & \textbf{\eer{1.1332488626616075}}   \\
   
    3D-Speaker (ERes2Net-large-lm) & \eer{0.52} & \eer{0.75} & \eer{1.44} \\
    
    ESPnet-SPK (SKA-WavLM) & \eer{0.516} & - & - \\ 
    WeSpeaker (ResNet-293) & \eer{0.532} & \eer{0.707} & \eer{1.311} \\
    
    \bottomrule
    \end{tabular}
    
}
    \caption{Comparison of EER on VoxCeleb1-O/E/H using the best-performing model from each toolkit. Kiwano achieves the lowest EER across all evaluation sets, highlighting its strong generalization and competitive performance among open-source frameworks.}  
    \label{tbl:other_toolkit}
\end{table}

\section{Conclusions}
\label{sec:conclusions}

In this paper, we introduced Kiwano, a modern, modular and extensible speaker verification toolkit designed for both research and production use. Kiwano integrates a wide range of SOTA architectures (e.g., ECAPA2, ReDimNet, Xi-Vector) and advanced back-end modules, achieving competitive results across multiple benchmarks. Its lightweight design, extensive recipe collection and reproducible pipelines make it a versatile toolkit for developing and evaluating speaker verification systems.

For the next release, we will focus on: 1) introducing new recipes for training models based on SSL encoders, enabling fine-tuning of SSL encoders for speaker embedding extraction; and 2) continuously integrating state-of-the-art (SOTA) speaker models, including novel network architectures and scoring back-ends.

\section{Acknowledgements}

This work was granted access to the HPC resources of IDRIS under the allocation AD011013257R4 made by GENCI.

\bibliographystyle{IEEEtran}
\bibliography{Odyssey2026_BibEntries}

@inproceedings{larcher2013alize,
  title={{ALIZE 3.0-open source toolkit for state-of-the-art speaker recognition}},
  author={Larcher, Anthony and Bonastre, Jean-Fran{\c{c}}ois and Fauve, Benoit and Lee, Kong Aik and L{\'e}vy, Christophe and Li, Haizhou and Mason, John and Parfait, Jean-Yves},
  year      = {2013},
  booktitle = {Interspeech},
  pages     = {2768--2772},
  doi       = {10.21437/Interspeech.2013-634},
  issn      = {2958-1796},
}

@inproceedings{fan2020cn,
  title={Cn-celeb: a challenging chinese speaker recognition dataset},
  author={Fan, Yue and Kang, JW and Li, LT and Li, KC and Chen, HL and Cheng, ST and Zhang, PY and Zhou, ZY and Cai, YQ and Wang, Dong},
  booktitle={International Conference on Acoustics, Speech and Signal Processing (ICASSP)},
  year={2020},
  organization={IEEE}
}

@inproceedings{paipuri2024ceems,
  title={{CEEMS: a resource manager agnostic energy and emissions monitoring stack}},
  author={Paipuri, Mahendra},
  booktitle={SC24-W: Workshops of the International Conference for High Performance Computing, Networking, Storage and Analysis},
  pages={1862--1866},
  year={2024},
  organization={IEEE}
}

@inproceedings{zheng2024score,
  title={Score calibration based on consistency measure factor for speaker verification},
  author={Zheng, Yu and Zhang, Yajun and Niu, Chuanying and Zhan, Yibin and Long, Yanhua and Xu, Dongxing},
  booktitle={International Conference on Acoustics, Speech and Signal Processing (ICASSP)},
  pages={12371--12375},
  year={2024},
  organization={IEEE}
}

@inproceedings{thienpondt2021idlab,
  title={{The idlab voxsrc-20 submission: Large margin fine-tuning and quality-aware score calibration in dnn based speaker verification}},
  author={Thienpondt, Jenthe and Desplanques, Brecht and Demuynck, Kris},
  booktitle={International Conference on Acoustics, Speech and Signal Processing (ICASSP)},
  pages={5814--5818},
  year={2021},
  organization={IEEE}
}

@inproceedings{Yakovlev_2024, 
   title={{Reshape Dimensions Network for Speaker Recognition}},
   url={http://dx.doi.org/10.21437/Interspeech.2024-2116},
   DOI={10.21437/interspeech.2024-2116},
   booktitle={Interspeech},
   publisher={ISCA},
   author={Yakovlev, Ivan and Makarov, Rostislav and Balykin, Andrei and Malov, Pavel and Okhotnikov, Anton and Torgashov, Nikita},
   year={2024},
   month=sep, pages={3235–3239},
}

@inproceedings{Park_2019,
   title={{SpecAugment: A Simple Data Augmentation Method for Automatic Speech Recognition}},
   url={http://dx.doi.org/10.21437/Interspeech.2019-2680},
   DOI={10.21437/interspeech.2019-2680},
   booktitle={Interspeech},
   publisher={ISCA},
   author={Park, Daniel S. and Chan, William and Zhang, Yu and Chiu, Chung-Cheng and Zoph, Barret and Cubuk, Ekin D. and Le, Quoc V.},
   year={2019},
   month=sep, pages={2613–2617},
    }

@article{snyder2015musan,
  title={{Musan: A music, speech, and noise corpus}},
  author={Snyder, David and Chen, Guoguo and Povey, Daniel},
  journal={arXiv preprint arXiv:1510.08484},
  year={2015},
  doi={10.48550/arXiv.1510.08484}
}

@inproceedings{yakovlev24_interspeech,
  title     = {{Reshape Dimensions Network for Speaker Recognition}},
  author    = {Ivan Yakovlev and Rostislav Makarov and Andrei Balykin and Pavel Malov and Anton Okhotnikov and Nikita Torgashov},
  year      = {2024},
  booktitle = {{Interspeech}},
  pages     = {3235--3239},
  doi       = {10.21437/Interspeech.2024-2116},
  issn      = {2958-1796},
}

@inproceedings{peng2023attention,
  title={An attention-based backend allowing efficient fine-tuning of transformer models for speaker verification},
  author={Peng, Junyi and Plchot, Old{\v{r}}ich and Stafylakis, Themos and Mo{\v{s}}ner, Ladislav and Burget, Luk{\'a}{\v{s}} and {\v{C}}ernock{\`y}, Jan},
  booktitle={2022 IEEE Spoken Language Technology Workshop (SLT)},
  pages={555--562},
  year={2023},
  organization={IEEE},
  doi={10.1109/SLT54892.2023.10022775}
}

@article{zheng20233d,
  title={{3d-speaker: A large-scale multi-device, multi-distance, and multi-dialect corpus for speech representation disentanglement}},
  author={Zheng, Siqi and Cheng, Luyao and Chen, Yafeng and Wang, Hui and Chen, Qian},
  journal={arXiv preprint arXiv:2306.15354},
  year={2023}
}

@inproceedings{lin2024voxblink,
  title={{Voxblink: A large scale speaker verification dataset on camera}},
  author={Lin, Yuke and Qin, Xiaoyi and Zhao, Guoqing and Cheng, Ming and Jiang, Ning and Wu, Haiying and Li, Ming},
  booktitle={International Conference on Acoustics, Speech and Signal Processing (ICASSP)},
  pages={10271--10275},
  year={2024},
  organization={IEEE}
}

@inproceedings{rouvier2022far,
  title={{Far-field speaker recognition benchmark derived from the DiPCo corpus}},
  author={Rouvier, Mickael and Mohammadamini, Mohammad},
  booktitle={Language Resources and Evaluation Conference (LREC)},
  pages={1955--1959},
  year={2022},
  url={https://aclanthology.org/2022.lrec-1.209/}
}

@article{lee2021xi,
  title={{Xi-vector embedding for speaker recognition}},
  author={Lee, Kong Aik and Wang, Qiongqiong and Koshinaka, Takafumi},
  journal={IEEE Signal Processing Letters},
  volume={28},
  pages={1385--1389},
  year={2021},
  publisher={IEEE},
  doi={10.1109/LSP.2021.3091932}
}

@article{dehak2010front,
  title={{Front-end factor analysis for speaker verification}},
  author={Dehak, Najim and Kenny, Patrick J and Dehak, R{\'e}da and Dumouchel, Pierre and Ouellet, Pierre},
  journal={IEEE Transactions on Audio, Speech, and Language Processing},
  volume={19},
  number={4},
  pages={788--798},
  year={2010},
  publisher={IEEE},
doi={10.1109/TASL.2010.2064307}
}

@inproceedings{yakovlev2023voxtube,
  title={{VoxTube: a multilingual speaker recognition dataset.}},
  author={Yakovlev, Ivan and Okhotnikov, Anton and Torgashov, Nikita and Makarov, Rostislav and Voevodin, Yuri and Simonchik, Konstantin},
  booktitle={Interspeech},
  pages={2238--2242},
  year={2023}
}

@article{hintz2024commonbench,
  title={{CommonBench: A larger scale speaker verification benchmark}},
  author={Hintz, Jan and Siegert, Ingo},
  journal={Symposium on Security and Privacy in Speech Communication (SPSC)},
  volume={2024},
  pages={17--20},
  year={2024}
}

@inproceedings{chung20_interspeech,
  title     = {{Spot the Conversation: Speaker Diarisation in the Wild}},
  author    = {Joon Son Chung and Jaesung Huh and Arsha Nagrani and Triantafyllos Afouras and Andrew Zisserman},
  year      = {2020},
  booktitle = {Interspeech},
  pages     = {299--303},
  doi       = {10.21437/Interspeech.2020-2337},
  issn      = {2958-1796},
}

@article{li2022cn,
  title={{Cn-celeb: multi-genre speaker recognition}},
  author={Li, Lantian and Liu, Ruiqi and Kang, Jiawen and Fan, Yue and Cui, Hao and Cai, Yunqi and Vipperla, Ravichander and Zheng, Thomas Fang and Wang, Dong},
  journal={Speech Communication},
  volume={137},
  pages={77--91},
  year={2022},
  publisher={Elsevier},
    doi={10.1016/j.specom.2022.01.002}
}

@inproceedings{wang2023wespeaker,
  title={{Wespeaker: A research and production oriented speaker embedding learning toolkit}},
  author={Wang, Hongji and Liang, Chengdong and Wang, Shuai and Chen, Zhengyang and Zhang, Binbin and Xiang, Xu and Deng, Yanlei and Qian, Yanmin},
  booktitle={International Conference on Acoustics, Speech and Signal Processing (ICASSP)},
  pages={1--5},
  year={2023},
  organization={IEEE},
  doi={10.1109/ICASSP49357.2023.10096626}
}

@article{sre2006nist,
  title={{NIST Speaker Recognition Evaluation}},
  author={SRE, NIST},
  year={2006}
}

@inproceedings{chung2018voxceleb2,
  title={{Voxceleb2: Deep speaker recognition}},
  author={Chung, Joon Son and Nagrani, Arsha and Zisserman, Andrew},
  booktitle = {Interspeech},
  pages     = {1086--1090},
  doi       = {10.21437/Interspeech.2018-1929},
  issn      = {2958-1796},
  year={2018}
}

@article{huh2024vox,
  title={{The VoxCeleb Speaker Recognition Challenge: A Retrospective}},
  author={Huh, Jaesung and Chung, Joon Son and Nagrani, Arsha and Brown, Andrew and Jung, Jee-weon and Garcia-Romero, Daniel and Zisserman, Andrew},
  journal={IEEE/ACM Transactions on Audio, Speech, and Language Processing},
  year={2024},
  publisher={IEEE},
doi={10.1109/TASLP.2024.3444456}
}

@inproceedings{jung2024espnet,
  title     = {{ESPnet-SPK: full pipeline speaker embedding toolkit with reproducible recipes, self-supervised front-ends, and off-the-shelf models}},
  author    = {Jee-weon Jung and Wangyou Zhang and Jiatong Shi and Zakaria Aldeneh and Takuya Higuchi and Alex Gichamba and Barry-John Theobald and Ahmed {Hussen Abdelaziz} and Shinji Watanabe},
  year      = {2024},
  booktitle = {{Interspeech}},
  pages     = {4278--4282},
  doi       = {10.21437/Interspeech.2024-1345},
  issn      = {2958-1796},
}

@inproceedings{thienpondt2023ecapa2,
  title={{Ecapa2: A hybrid neural network architecture and training strategy for robust speaker embeddings}},
  author={Thienpondt, Jenthe and Demuynck, Kris},
  booktitle={Automatic Speech Recognition and Understanding Workshop (ASRU)},
  pages={1--8},
  year={2023},
  organization={IEEE}
}

@inproceedings{snyder2018x,
  title={{X-vectors: Robust dnn embeddings for speaker recognition}},
  author={Snyder, David and Garcia-Romero, Daniel and Sell, Gregory and Povey, Daniel and Khudanpur, Sanjeev},
  booktitle={International Conference on Acoustics, Speech and Signal Processing (ICASSP)},
  pages={5329--5333},
  year={2018},
  organization={IEEE},
  doi={10.1109/ICASSP.2018.8461375}
}

@inproceedings{povey2011kaldi,
  title={{The Kaldi speech recognition toolkit}},
  author={Povey, Daniel and Ghoshal, Arnab and Boulianne, Gilles and Burget, Lukas and Glembek, Ondrej and Goel, Nagendra and Hannemann, Mirko and Motlicek, Petr and Qian, Yanmin and Schwarz, Petr and others},
  booktitle={IEEE 2011 workshop on automatic speech recognition and understanding},
  volume={1},
  pages={5--1},
  year={2011},
  organization={Hawaii}
}

@article{ravanelli2021speechbrain,
  title={{Open-source conversational ai with speechbrain 1.0}},
  author={Ravanelli, Mirco and Parcollet, Titouan and Moumen, Adel and De Langen, Sylvain and Subakan, Cem and Plantinga, Peter and Wang, Yingzhi and Mousavi, Pooneh and Della Libera, Luca and Ploujnikov, Artem and others},
  journal={Journal of Machine Learning Research},
  volume={25},
  number={333},
  pages={1--11},
  year={2024},
  doi={doi/10.5555/3722577.3722910}
}

@inproceedings{rouvier2021studying,
  title={{Studying squeeze-and-excitation used in CNN for speaker verification}},
  author={Rouvier, Mickael and Bousquet, Pierre-Michel},
  booktitle={Automatic Speech Recognition and Understanding Workshop (ASRU)},
  pages={1110--1115},
  year={2021},
  organization={IEEE}
}

@inproceedings{bousquet2023jeffreys,
  title={{Jeffreys divergence-based regularization of neural network output distribution applied to speaker recognition}},
  author={Bousquet, Pierre-Michel and Rouvier, Mickael},
  booktitle={International Conference on Acoustics, Speech and Signal Processing (ICASSP)},
  pages={1--5},
  year={2023},
  organization={IEEE}
}

@INPROCEEDINGS{Alam18,
  author = {Jahangir Alam and Gautam Bhattacharya and Patrick Kenny},
  title = {{Speaker Verification in Mismatched Conditions with Frustratingly Easy Domain Adaptation}},
  booktitle = {Odyssey 2018},
    doi={10.21437/Odyssey.2018-25},
year={2018},
  url={http://dx.doi.org/10.21437/Odyssey.2018-25}
}

@article{Lee18,
  author    = {Kong Aik Lee and
               Qiongqiong Wang and
               Takafumi Koshinaka},
  title     = {{The CORAL+ Algorithm for Unsupervised Domain Adaptation of PLDA}},
  journal   = {CoRR},
  volume    = {abs/1812.10260},
  year      = {2018},
  url       = {http://arxiv.org/abs/1812.10260},
  archivePrefix = {arXiv},
  eprint    = {1812.10260},
  bibsource = {dblp computer science bibliography, https://dblp.org}
}

@inproceedings{Bousquet2019,
  author={Pierre-Michel Bousquet and Mickael Rouvier},
  title={{On Robustness of Unsupervised Domain Adaptation for Speaker Recognition}},
  year=2019,
  booktitle={Interspeech},
  pages={2958--2962},
  doi={10.21437/Interspeech.2019-1524},
  url={http://dx.doi.org/10.21437/Interspeech.2019-1524}
}

@inproceedings{karam2011towards,
  title={Towards reduced false-alarms using cohorts},
  author={Karam, Zahi N and Campbell, William M and Dehak, Najim},
  booktitle={International Conference on Acoustics, Speech and Signal Processing (ICASSP)},
  pages={4512--4515},
  year={2011},
  organization={IEEE}
}

@inproceedings{cumani2023adaptive,
  title={From adaptive score normalization to adaptive data normalization for speaker verification systems.},
  author={Cumani, Sandro and Sarni, Salvatore and others},
  booktitle = {Interspeech},
  pages={5296--5300},
  year={2023}
}

@inproceedings{shum2010unsupervised,
  title={Unsupervised speaker adaptation based on the cosine similarity for text-independent speaker verification},
  author={Shum, Stephen and Dehak, Najim and Dehak, Reda and Glass, James},
  booktitle={Odyssey 2010},
  pages={paper--16},
  year={2010}
}

\end{document}